\documentclass[12pt,a4paper]{article}
\usepackage{amsmath}
\usepackage[dvips]{graphics}

\newcommand{\bea}{\begin{eqnarray*}}
\newcommand{\eea}{\end{eqnarray*}}

\newcommand{\bean}{\begin{eqnarray}}
\newcommand{\eean}{\end{eqnarray}}

\newcommand{\eps}{\epsilon}

\begin{document}

\title{The Penrose inequality and apparent horizons}

\author{Ishai Ben-Dov\thanks{Electronic mail: \tt ibd@midway.uchicago.edu} 
\\\it Enrico Fermi Institute and Department of Physics 
\\ \it University of Chicago \\ \it 5640 S. Ellis Avenue    
\\ \it Chicago, Illinois 60637-1433}
\maketitle

\begin{abstract}

A spherically symmetric spacetime is presented with an initial data set that is
asymptotically flat, satisfies the dominant energy condition, and such that on
this initial data $M<\sqrt{A/16\pi}$, where $M$ is the total (ADM) mass and
$A$ is the area of the apparent horizon. This provides a counterexample to a
commonly stated version of the Penrose inequality, though it does not
contradict the ``true'' Penrose inequality.

\end{abstract}


\section{Introduction}   \label{introduction}

For over 30 years, the Penrose inequality has been a major open question in
classical general relativity, closely tied to an even bigger open question,
the cosmic censorship conjecture. In 1973 Penrose \cite{Penrose1} discussed
attempts to find violations of the cosmic censorship conjecture.
These attempts led him to formulate a certain inequality, the derivation of
which relies so heavily on cosmic censorship that a violation of this
inequality would go a long way towards contradicting it.
Likewise, a proof of this inequality would strengthen the common belief in
the validity of the cosmic censorship conjecture, though by itself a proof of
the inequality would, of course, not serve as an actual proof of the
conjecture.

Penrose's original scenario was of a collapsing shell of null dust. A slightly
modified presentation was given by Jang and Wald \cite{JangWald} who discussed
this in the context of Cauchy initial data, i.e.\ a 3-manifold with a
(complete) Riemannian metric, a symmetric tensor field representing the
extrinsic curvature, and matter density and current. Repeating their
discussion, the argument goes as follows:
consider asymptotically flat initial data with Arnowitt--Deser--Misner (ADM)
mass $M$ and an event horizon of area $A_E$. If cosmic censorship holds, this
spacetime is expected to settle, eventually, to a Kerr--Newman black
hole solution. Let $M_K$ be its mass and $A_K$ be the area of its event
horizon. For the Kerr solution $M_K \geq \sqrt{A_K/16\pi}$. By Hawking's area 
theorem \cite{Hawking1}, which assumes cosmic censorship, the area of the 
event horizon is non-decreasing, hence $A_E \leq A_K$. Furthermore the 
Bondi--Sachs energy radiated to null infinity is positive, so that the mass is 
non-increasing, $M \geq M_K$. It follows that for the initial data, $M$ and
$A_E$ should satisfy
\bean
M \, \geq \, M_K \, \geq \, \sqrt{A_K/16\pi} \, \geq \, \sqrt {A_E/16\pi}
\eean

Assuming cosmic censorship and reasonable energy conditions the
apparent horizon lies within the event horizon \cite{WaldGR}. This leads to:

\noindent
{\bf The Penrose inequality}: Given any asymptotically flat initial data
satisfying the dominant energy condition then
\bean \label{PenroseIneq}
M \, \geq \, \sqrt{\mathcal{A}/16\pi}
\eean

\noindent
with $M$ the total mass and $\mathcal{A}$ the minimum area required to enclose
the apparent horizon.

\medskip
Notice this rather delicate statement: though the apparent horizon lies inside
the region enclosed by the event horizon, it need not be true that the area of
the apparent horizon is smaller than the area of the event horizon. However,
$\mathcal{A} \leq A_E$ still holds, hence the form that the Penrose inequality
takes.

In recent years there has been much progress in proving the Penrose inequality
in certain cases, though the general case is still open. Of note are proofs of
the Riemannian Penrose inequality---the Penrose inequality in the time
symmetric case. In this case the initial data has vanishing extrinsic
curvature everywhere.\footnote{As a result, as to satisfy the initial value
constraints, it follows that the matter current must vanish as well.}\,
The requirement that the dominant energy condition be satisfied
implies, in the time symmetric case, that the scalar curvature of this
3-manifold is everywhere non-negative.\footnote{In fact the weak energy
condition implies the same thing. The Penrose inequality in the time symmetric
case has therefore been proven with either energy condition.}\,
Furthermore, the vanishing of the extrinsic
curvature implies that the apparent horizon is an outermost minimal surface
\cite{BrayChrusc}. Hence in this case, $\mathcal{A}$ is equal to $A$, the area
of the apparent horizon. Thus, the Riemannian Penrose inequality 
\cite{BrayChrusc} is defined to be:

\noindent
{\bf The Riemannian Penrose inequality}: Given any asymptotically flat initial
data satisfying the dominant energy condition, if, in addition, the
extrinsic curvature vanishes, then
\bean
M \, \geq \, \sqrt{A/16\pi}
\eean

\noindent
with $M$ the total mass and $A$ the area of the apparent horizon.

\medskip

As mentioned, this has been proven by Huisken and Ilmanen \cite{HuiIlm},
in the case where the apparent horizon is connected, thereby making rigorous
the argument of \cite{JangWald} based on an original idea of
Geroch \cite{Geroch}. More recently, Bray \cite{Bray} proved
this in a more general case, when the apparent horizon may consist of 
several disconnected components.

\bigskip

The form of the Riemannian Penrose inequality suggests a possible
generalization:

\noindent
{\bf The apparent horizon Penrose inequality}: Given any asymptotically flat
initial data satisfying the dominant energy condition, then
\bean
M \, \geq \, \sqrt{A/16\pi}
\eean

\noindent
with $M$ the total mass and $A$ the area of the apparent horizon. (This is 
simply a modification of the Riemannian Penrose inequality by relaxing the
requirement of time symmetry.)

\medskip

The time symmetric case is a special one, as there, as mentioned above, one can
replace the minimum area required to enclose the apparent horizon with the area
of the apparent horizon itself. Thus Penrose's original reasoning still applies
to that case. But without the extra assumption of time symmetry, there does not
seem to be any physical reason to expect the apparent horizon Penrose
inequality to hold in general. Nevertheless there have been numerous
appearances of this conjecture in the literature (e.g.\ \cite{Malec&02,
Szabad,AshKrish1,Karkow&93,Frauen}). With regard to the apparent horizon
Penrose inequality, Bray and Chrusciel \cite{BrayChrusc} state that a 
counterexample would not be terribly surprising, although it would be very 
interesting.

The main purpose of the present work is to provide an explicit counterexample
to the apparent horizon Penrose inequality.

\bigskip

The remainder of the paper is organized as follows:
Section~\ref{TrappedSurfaces}
covers the required background on outer/inner trapped surfaces, especially in 
the Schwarzschild and Robertson--Walker (RW) spacetimes.
In Section~\ref{OpSnyST} spacetimes very much like (and including) the
Oppenheimer--Snyder (OS) collapse
model are described. These are used in Section~\ref{Counterexample} to obtain a
counterexample to the apparent horizon Penrose inequality. A relevant issue
pertaining to certain quasi-local constructions is discussed
in Section~\ref{QuasiLocal}. The paper concludes with a discussion of various
versions of the inequality in an attempt to clarify what has been proven and
which versions still stand a chance of holding.

\medskip

The notation and conventions follow Wald \cite{WaldGR}.


\section{Trapped surfaces} \label{TrappedSurfaces}

In order to make precise the inequality in question, the notion of
outer/inner trapped surfaces is required.

Given a closed spacelike 2-surface, at each point there are precisely two 
future-directed null rays normal to the surface at that point. If the spacetime
is asymptotically flat and the 2-surface is the boundary of a region that does
not extend to the asymptotically flat end, one can distinguish between
outgoing (towards the asymptotically flat region) and ingoing null rays.
This motivates the next definitions, closely following Wald \cite{WaldGR}.

On a Cauchy surface, an {\it outer} (resp.\ {\it inner}) {\it trapped surface}
is a compact smooth spacelike 2-manifold, which is the boundary of a region
that does not extend to the asymptotically flat end, such that the expansion
of outgoing (resp.\ ingoing) future-directed null geodesics normal to it is
everywhere negative. A surface which is both outer trapped and inner trapped
is a {\it trapped surface}.\footnote{In this case, one can
discard the requirement of it being a boundary, since if both expansions are
negative, it does not matter which one is inner and which one is outer, as in
any case the surface is trapped in both directions.}\, A 
{\it marginally outer} (resp.\ {\it inner}) {\it trapped surface} is one where
the above requirements are weakened so as to demand that the expansion of
outgoing (resp.\ ingoing) null geodesics be only non-positive.
On a Cauchy surface a {\it trapped region} is a region that does not extend to
the asymptotically flat end and whose boundary is a marginally outer trapped
surface. Finally, the {\it total trapped region} on a Cauchy surface is the
closure of the union of all trapped regions in the Cauchy surface.

A trapped surface is an indication of a strong gravitational field. When the
gravitational field is not so strong, one normally expects that the ingoing
null rays will be converging while the outgoing ones will be diverging. Hence,
for example, in Minkowski spacetime every 2-sphere with fixed $r$ and 
$t$ (in spherical coordinates) is inner trapped but the expansion of outgoing 
null geodesics in this case is everywhere positive (i.e.\ they diverge).
In the other extreme, 2-spheres inside a Schwarzschild black hole region are
trapped.

Given initial data, one can search the initial data 3-manifold for trapped
surfaces. The notion of apparent horizon is extremely useful in this context,
as the {\it apparent horizon} is defined to be the boundary of the
total trapped region in the initial data 3-manifold.
As a result the apparent horizon is marginally outer trapped with the expansion
of future-directed, outgoing, null geodesics normal to it everywhere vanishing.
Spherically symmetric initial data implies that the total trapped region is
also spherically symmetric and therefore the apparent horizon is spherically
symmetric as well.

It is known that in an asymptotically predictable spacetime satisfying the 
null energy condition, any marginally outer trapped surface 
(specifically, the apparent horizon) lies inside a black hole. Hence apparent 
horizons, which are local and therefore relatively easy to locate, are
indicative of black holes. In general the black hole's event horizon does not
coincide with the apparent horizon. But being global, the event horizon
requires a full knowledge of the spacetime's causal structure before the event
horizon's actual existence is known, not to mention its exact location.
Hence, for various applications, apparent horizons are more immediately
accessible and practical.

With these definitions it is worthwhile to explore two specific spacetimes for 
the location of trapped surfaces. The discussion will prove useful, not only as
concrete examples of the definitions above, but more importantly since these 
spacetimes will be ingredients in later constructions.

\subsection{Maximally extended Schwarzschild spacetime}

\begin{figure}
      \begin{center}
    \resizebox{13cm}{!}{\includegraphics{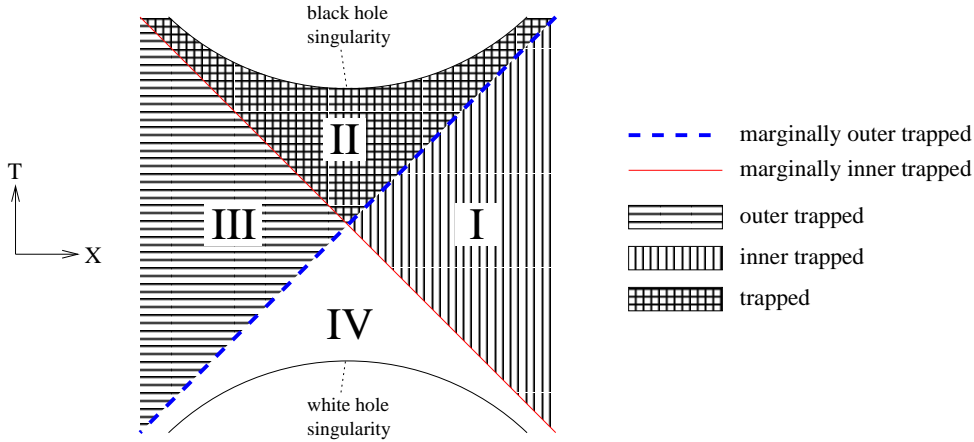}}
  \end{center}
  \caption{A spacetime diagram of the Kruskal extension of a Schwarzschild 
spacetime. Two angular dimensions are suppressed. Each point therefore 
represents a 2-sphere.}
  \label{KruskalFig}
\end{figure}

A spacetime diagram of the Kruskal extension of the Schwarzschild spacetime is
shown in Fig.\ \ref{KruskalFig}. In this case there are two asymptotically flat
regions (Regions I and III), hence a designation of one of these regions as
representing ``infinity'' is required in order to discuss outer and inner
trapped surfaces. From now on, Region I is taken to contain infinity.
The event horizon with respect to this choice of infinity is the surface (in
the figure, a curve)
separating Region I from II and separating Region III from IV. 
(From the perspective of the other asymptotically flat region, i.e.\ taking
Region III to contain infinity, the figure is reversed (left to right) and so
the event horizon in that case would be the surface separating Region II
from III, and I from IV.)\, The 2-spheres in Regions II and III are outer
trapped. The 2-spheres in Regions
I and II are inner trapped. Hence Region II consists of trapped 2-spheres. 
Region IV consists of 2-spheres that are neither inner nor outer trapped
(it represents the white hole
region and the 2-spheres in it would be outer and inner trapped if in the
definitions above, future-directed null geodesics were replaced by
past-directed ones). On any spherically symmetric Cauchy surface, the apparent
horizon coincides with the event horizon. On it, the expansion of 
the outgoing null geodesics vanishes and the expansion of the ingoing ones is
negative after the moment of time symmetry (in the portion separating Regions I
and II) and is positive before that (in the portion separating Regions III
and IV). The change in sign of the expansion of ingoing null geodesics is
related to the difference between the white hole and black hole regions.
This will be crucial to the construction of the counterexample.

\subsection{Closed, dust-filled Robertson--Walker spacetime}

A closed RW spacetime is a homogeneous, isotropic spacetime that starts from a
singularity (the big bang) and ends in another (the big crunch). It has the
line element
\bean \label{closedRWmetric}
d s^2=-d \tau^2 + a^2(\tau)\Big(d\chi^2 + \sin^2\chi d\Omega^2 \Big)
\eean

\noindent
with $\chi\in[0,\pi]$. In the case of pressureless matter (dust) it is useful
to introduce a parameter $\eta$ in terms of which the scale factor $a$ and
proper time $\tau$ take the following forms \cite{MTW}:
\bean
a(\eta) = \frac{1}{2} a_m(1-\cos(\eta)) \label{cyclic1} \\
\tau(\eta) = \frac{1}{2} a_m(\eta - \sin(\eta)) \label{cyclic2}
\eean

\noindent
where $\eta \in [0,2\pi]$ for the full evolution from big bang to big crunch
and $a_m$ is the value of the scale factor at the moment of maximum expansion.

As this RW spacetime is closed, there is no a priori way to assign a notion of
outer and inner. Instead, given the line element above and the specific slice
to be described later, it will be useful to focus on the north and south poles.
Moreover, as the spacetimes and initial data later discussed are all
spherically symmetric, then, since in this case the apparent horizon is
spherically symmetric as well, the discussion can be limited to 2-spheres of
constant $\chi$ and $\tau$. The following definitions are limited to closed
RW spacetimes.

The north pole is the point $\chi=0$ and so let a {\it north trapped surface}
denote a 2-sphere of constant $\chi$ and $\tau$ such that the expansion of
future, north-pole-directed (i.e.\ oriented towards decreasing $\chi$)
null geodesics normal to it is everywhere negative.
A {\it marginally north trapped surface} denotes weakening the last requirement
so that the expansion is only required to be non-positive.
A {\it south trapped surface} and a {\it marginally south trapped surface}
are defined similarly. In the context of a closed RW spacetime, a trapped
surface is a 2-sphere of constant $\chi$ and $\tau$ that is both north trapped
and south trapped.
Later, when a portion of such a spacetime will be attached to another in
forming a new spacetime with an asymptotically flat region, the north/south
ends will be assigned an inner/outer interpretation. (For consistency, 
in all such constructions the south end will be the outer one and the north
end the inner one. In this case a south trapped surface is also an outer
trapped surface, and a north trapped surface is also an inner trapped one.)

\begin{figure} [!t]
      \begin{center}
    \resizebox{11cm}{!}{\includegraphics{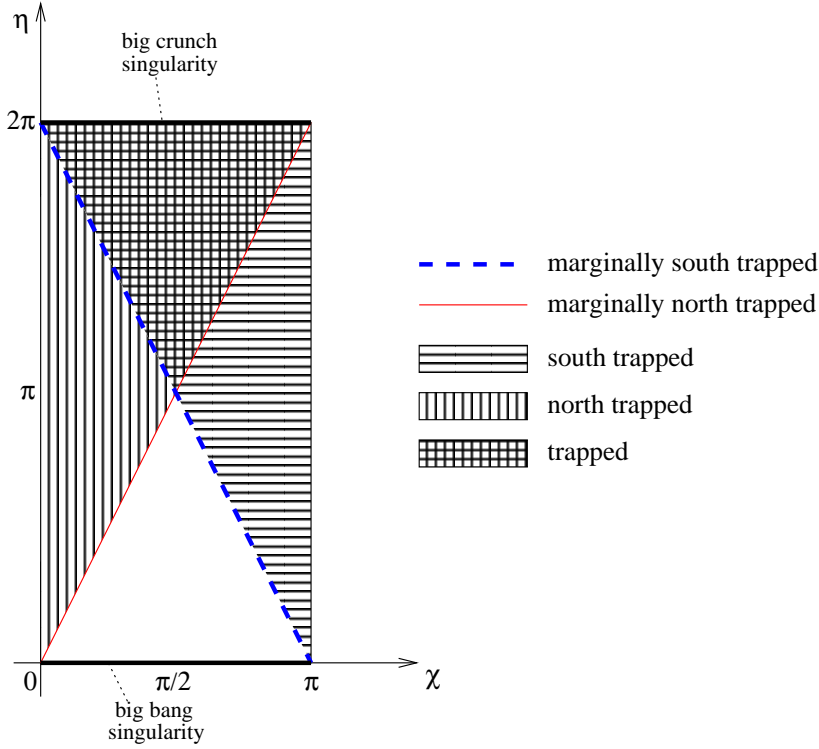}}
  \end{center}
  \caption{A spacetime diagram of a closed, dust, RW spacetime.}
  \label{RWFig}
\end{figure}

With the line element above, the tangents to the future, radial,
south-directed, null geodesics are given by\footnote{The notation is consistent
with later letting $l^a$ be tangent to outgoing geodesics and $n^a$ tangent to
ingoing ones.} 
\bean
l^a = \frac{1}{\sqrt{2}} \Bigg[ \Big(\frac{\partial}{\partial \tau}\Big)^a + 
\frac{1}{a(\tau)} \Big(\frac{\partial}{\partial \chi}\Big)^a \Bigg]
\eean

\noindent
and the tangents to the north-directed ones are given by
\bean
n^a = \frac{1}{\sqrt{2}} \Bigg[ \Big(\frac{\partial}{\partial \tau}\Big)^a - 
\frac{1}{a(\tau)} \Big(\frac{\partial}{\partial \chi}\Big)^a \Bigg]
\eean

\noindent
with $l^a n_a = -1$.

Given a radial null vector field $k^a$ its expansion is given by
\bean
\theta_{(k)} = {q^a}_b \nabla_a k^b
\eean

\noindent
where $q_{ab} \equiv r^2\Big((d\theta)_a (d\theta)_b + 
\sin^2\theta\, (d\phi)_a (d\phi)_b\Big)$ is the metric on the 2-sphere. A
simple calculation yields:
\bean
\theta_{(l)} = \frac{\sqrt{2}}{a}\big(\dot{a}+\cot \chi) \quad\quad
\theta_{(n)} = \frac{\sqrt{2}}{a}\big(\dot{a}-\cot \chi)
\eean

\noindent
where $\dot{a} \equiv \frac{d a}{d \tau}$. Using (\ref{cyclic1}), 
(\ref{cyclic2}), these can be used to solve for the surfaces
$\chi_{(l)}(\eta)$ and $\chi_{(n)}(\eta)$ where the corresponding expansions of
null geodesics vanish. The result is:
\bean
\chi_{(l)}(\eta) = \pi - \frac{\eta}{2} \quad\quad
\chi_{(n)}(\eta) = \frac{\eta}{2} \label{ChiOfEta}
\eean

\noindent
Let the 3-surfaces defined by $\chi(\eta) = \chi_{(l)}(\eta)$ and
$\chi(\eta) = \chi_{(n)}(\eta)$ be denoted $\Sigma_{(l)}$ and $\Sigma_{(n)}$
respectively.

A natural question, which, as will be discussed later, is relevant to the
construction of the counterexample and further discussions,
is whether $\Sigma_{(l)}$ and $\Sigma_{(n)}$ are timelike, null,
or spacelike. This is answered in the appendix, where the following result
is shown:

\vskip.3cm
\noindent
Given a RW spacetime of any spatial geometry (closed,
flat, or open) and matter satisfying an equation of state of the form $P = 
\omega \rho$ with $\rho>0$, then the type of
$\Sigma_{(l)}$ and $\Sigma_{(n)}$\footnote{In RW spacetimes of any spatial
geometry, $\Sigma_{(l)}$ and $\Sigma_{(n)}$ are defined in a similar way,
in terms of surfaces where the expansions towards
increasing and decreasing radial coordinate vanish.} depends on $\omega$ alone:
\bea
\Sigma_{(l)} \text{ and } \Sigma_{(n)} \text{ are }
\left\{ \begin{array}{l@{\quad\quad\quad}l}
\text{timelike,} & -1< \omega < \frac{1}{3} \\ 
\text{null,} & \omega = \frac{1}{3} , -1 \\
\text{spacelike,} & \omega > \frac{1}{3} \text{ or } \omega < -1
\end{array} \right.
\eea

\vskip.3cm
\noindent
This implies that in the case of dust ($\omega=0$) the surfaces are
timelike.\footnote{As it is known that in the
closed dust RW spacetime a radial null ray can go all around the universe
(i.e.\ from $\chi=0$ to $\chi=\pi$ and then back to $\chi=0$) during the
entire evolution of the spacetime, it seems reasonable that $\Sigma_{(l)}$,
$\Sigma_{(n)}$ are therefore timelike in this case (as they cover 
only half the distance in the same time).}\, A spacetime diagram of a closed,
dust-filled RW spacetime is shown in Fig.\ \ref{RWFig}.


\section{Oppenheimer--Snyder-like spacetimes} \label{OpSnyST}

In 1939, Oppenheimer and Snyder \cite{OppSny} gave the first example of a
spacetime that describes the gravitational collapse of matter into a black
hole. In their work the matter was a ball of dust (i.e.\ pressureless,
homogeneous fluid), as a model for the gravitational collapse of a 
neutron star.

One way of constructing new solutions to Einstein's field equations from known
solutions is by matching two portions of two different spacetimes together.
The OS-like spacetimes are obtained by matching a part of a dust-filled RW
spacetime with a part of a Schwarzschild spacetime. The Israel--Darmois
junction conditions \cite{Israel1,Darmois} are conditions to be
imposed on the matching surface, the surface where the two portions are joined,
in a way to obtain such a solution. Only a specific case of such matchings will
be used---a matching with no extra matter between the two parts matched
(i.e.\ no thin shell of matter) and one where the matching surface is timelike.
In this case the junction conditions are equivalent to the 3-metric on the
matching surface being the same when evaluated in either portion being matched,
and similarly for the extrinsic curvature of this surface in each portion.

Although OS-like spacetimes can be constructed using RW
spacetimes of any spatial geometry (closed, flat, or open), the current
discussion will be limited to closed RW spacetimes, as these will suffice for
the construction of the counterexample and discussion thereafter.

Consider first the original OS model. The discussion follows \cite{MTW}. (An
excellent treatment of the original OS model can be found
in \cite{Poisson}.)\, Two spacetimes are taken: a closed, dust-filled RW 
spacetime with a line element given by (\ref{closedRWmetric}) and a
Schwarzschild spacetime with the line element
\bean
d s^2= -\Big(1-\frac{2M}{R}\Big)d t^2 + \Big(1-\frac{2M}{R}\Big)^{-1} d R^2 + 
R^2 d\Omega^2
\eean

\begin{figure}
      \begin{center}
    \resizebox{10cm}{!}{\includegraphics{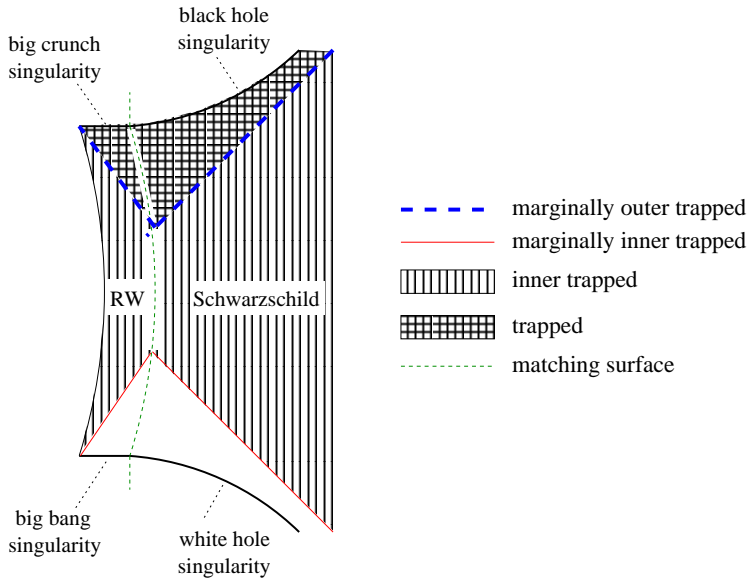}}
  \end{center}
  \caption{A spacetime diagram of the original OS model.}
  \label{OppSnyFig}
\end{figure}

The RW spacetime is cut at some fixed coordinate $\chi$,
$\chi_0 \leq \frac{\pi}{2}$, and only the northern portion, $\chi \leq \chi_0$
is kept. The Schwarzschild spacetime is cut along a surface spanned by
radial timelike geodesics at rest at $t=0$ at Schwarzschild coordinate radius
$R_0 \ge 2M$ in Region I of Fig.\ \ref{KruskalFig}. The portion
that contains the asymptotically flat end of Region I is kept. The two portions
kept (of RW and Schwarzschild) are pasted together along the surfaces
originally used to cut the two spacetimes.

As before, let $a_m$ be the maximum value attained by the scale factor $a$. The
junction conditions are satisfied and a new spacetime is obtained
when~\cite{MTW}
\bean
R_0=a_m \sin\chi_0 \label{junction1} \\
M=\frac{1}{2}a_m \sin^3\chi_0 \label{junction2}
\eean

The resulting spacetime includes a white hole singularity continuously
joined to a big bang singularity. The ball of matter starts expanding as matter
comes out from the white hole. The $t=0$ hypersurface of Schwarzschild,
coincides with the maximum expansion of the ball, a moment of time symmetry
in the RW region. Finally the ball of
matter collapses and a black hole is formed ending in a big crunch singularity
continuously joined to a black hole singularity. A spacetime diagram is shown
in Fig.\ \ref{OppSnyFig}.

As the matching is done with no thin matter shell, i.e.\ the extrinsic 
curvature is the same when evaluated in both portions matched, then the
marginally outer/inner trapped surfaces are continuous 
as can be seen in the spacetime diagram. Note how in this matching the
marginally outer (resp.\ inner) trapped surfaces in Fig.\ \ref{OppSnyFig} in
the RW portion are just the marginally south (resp.\ north) trapped
surfaces in Fig.\ \ref{RWFig}.

A 2-parameter family of solutions is obtained in this manner. One parameter is
the total mass of the spacetime, as this can be rescaled by taking 
$M \to \alpha M$ as well as $\rho \to \alpha \rho$ and $R \to \alpha R$, 
where $\rho = T_{ab} u^a u^b$ is the density of matter in the RW region. 
The other parameter is $\chi_0$, or equivalently, given a total mass $M$, at
what value of $R_0$ the Schwarzschild spacetime is cut. The latter freedom
will play an important role in the construction of the counterexample.

\begin{figure}
      \begin{center}
    \resizebox{14cm}{!}{\includegraphics{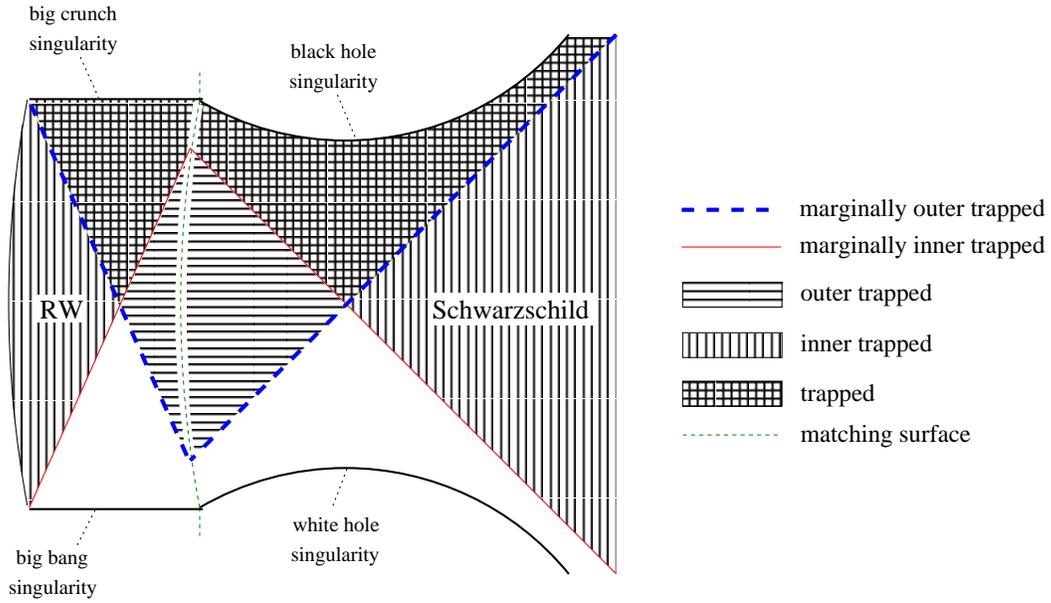}}
  \end{center}
  \caption{A spacetime diagram of a generalized OS spacetime with
    $\chi_0 > \frac{\pi}{2}$.}
  \label{OppSnyExtFig}

\end{figure}

\begin{figure}
      \begin{center}
    \resizebox{11cm}{!}{\includegraphics{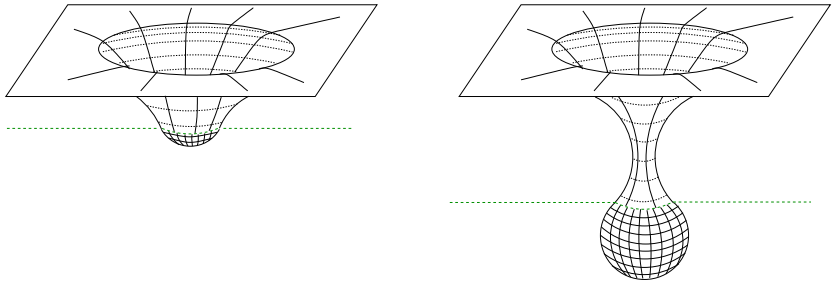}}
  \end{center}
  \caption{Embedding diagrams of OS-like spacetimes sliced at the moment 
    of time symmetry. One angular dimension is suppressed by setting
    $\theta~=~\pi/2$. Left: a normal OS spacetime. Right: a generalized OS
    spacetime with $\chi_0~>~\frac{\pi}{2}$.}
  \label{OppSnyEmbedFig}
\end{figure}

The construction above can be generalized by dropping the requirement
$\chi_0 \leq \frac{\pi}{2}$. This adds spacetimes with a ``star inside
the black hole'', i.e.\ the dust RW region never gets to Region I of the
Schwarzschild spacetime. Instead the surface of matching passes through the
other side of the Schwarzschild wormhole, Region III. This corresponds to
taking $\chi_0 > \frac{\pi}{2}$ as well as cutting the Schwarzschild spacetime
using geodesics that at the moment of time symmetry are located at
Schwarzschild coordinate radius $R_0$ (with $R_0>2M$) {\bf in Region III}
of Fig.\ \ref{KruskalFig}. 
The portion to be kept in each spacetime is the same, i.e.\ keeping the portion
$\chi \leq \chi_0$ in the RW spacetime and keeping the portion of
Schwarzschild that contains the asymptotically flat end in Region~I. The
junction conditions are unchanged as well and take the form (\ref{junction1}),
(\ref{junction2}).

A spacetime diagram of the resulting spacetime is shown in Fig.\ 
\ref{OppSnyExtFig}. An embedding diagram of both a normal OS spacetime and a
generalized one with $\chi_0 > \frac{\pi}{2}$ is shown in
Fig.\ \ref{OppSnyEmbedFig}. 


\section{A counterexample to the apparent horizon Penrose
inequality} \label{Counterexample}

The counterexample is obtained by taking a portion of a generalized OS
spacetime and matching it further. After the RW portion of this generalized OS
spacetime comes a second Schwarzschild region, with a bigger mass parameter.
It is then possible to slice this spacetime with a spacelike surface and get
initial data that provides a counterexample. A spacetime diagram of such a
construction is shown in Fig.\ \ref{CounterFig} and the precise details are now
described.

\medskip

\noindent
The construction is obtained by matching the following portions:

\begin{enumerate}

\item{A Schwarzschild spacetime with mass $M$ is cut along a surface
spanned by radial timelike geodesics at rest at $t=0$ at Schwarzschild
coordinate radius $R_0 > 2M$ in Region III of Fig.\ \ref{KruskalFig}. The
portion that contains the asymptotically flat end of Region I is kept. This
is the right Schwarzschild region in Fig.\ \ref{CounterFig}.}

\item{A closed, dust-filled RW spacetime with $a_m$ being the maximum value
attained by its scale factor. This spacetime is cut along two surfaces of
fixed $\chi$, i.e.\ $\chi=\chi_0$ and $\chi=\chi_1$, with
$\frac{\pi}{2} \leq \chi_1 < \chi_0 < \pi$
(thus dividing this spacetime into 3 separate
regions). The portion kept for the purposes of matching is the region
$\chi_1 \leq \chi \leq \chi_0$. This is the right RW region in
Fig.\ \ref{CounterFig}.}

\item{A second Schwarzschild spacetime, this time with mass $M_1$, is
cut along 2 surfaces (dividing it as well into 3 separate regions).
The first cut is along a surface spanned by radial timelike geodesics at
rest at $t=0$ at Schwarzschild coordinate radius $R_1 \ge 2M_1$ in Region III
of Fig.\ \ref{KruskalFig}, and the second cut is along another surface
spanned by radial timelike geodesics that are at rest at $t=0$ at
Schwarzschild coordinate radius $R_2 > R_1$ in
Region III of Fig.\ \ref{KruskalFig}. The portion kept is the middle one.
This is the left Schwarzschild region in Fig.\ \ref{CounterFig}.}

\item{A second closed, dust-filled RW spacetime, this time with
$a^\prime_m$ being the maximum value attained by its scale factor. This
spacetime is cut along a surface of fixed $\chi$, $\ \frac{\pi}{2} < \chi_2
< \pi$ keeping the portion $\chi \leq \chi_2$.
This is the left RW region in Fig.\ \ref{CounterFig}.}

\end{enumerate}

\medskip
The portions are matched together in consecutive order, i.e.\ 1-2-3-4, as can 
be seen in Fig.\ \ref{CounterFig}. The exact details of these matchings are:
\medskip

\begin{figure}
      \begin{center}
    \resizebox{13.8cm}{!}{\includegraphics{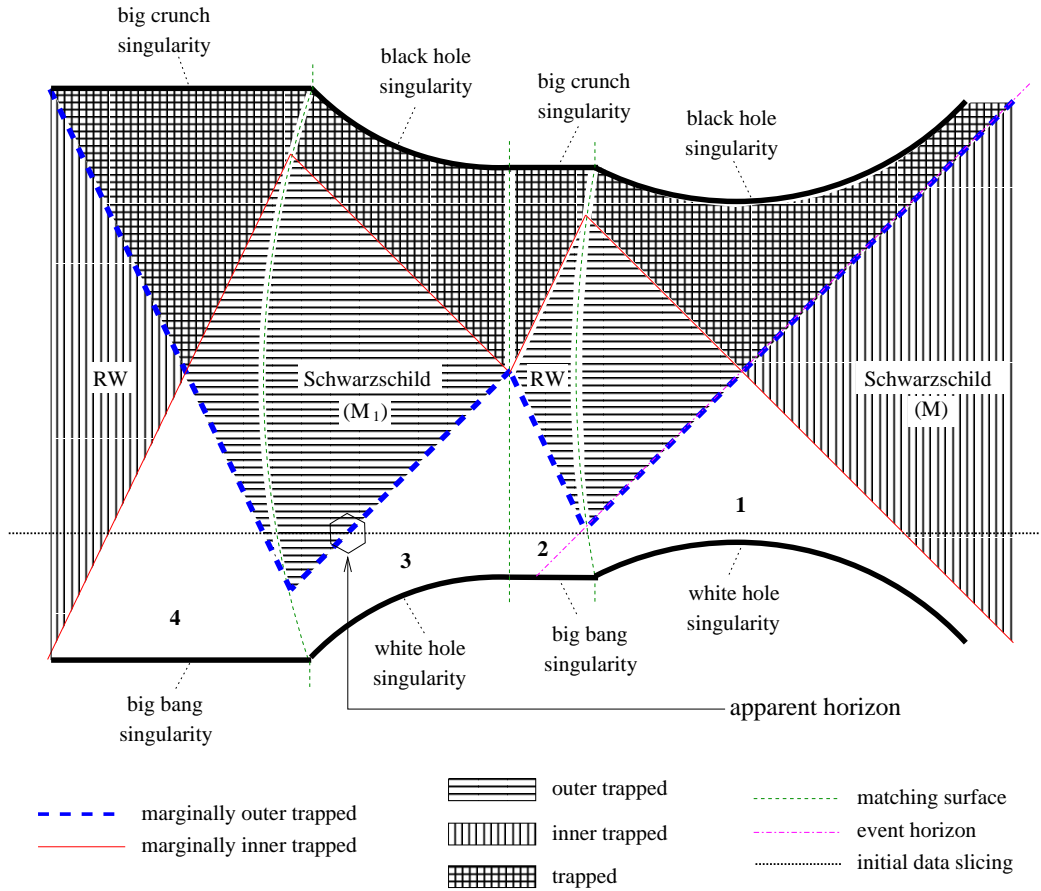}}
  \end{center}
  \caption{A spacetime diagram of a counterexample to the apparent horizon
    Penrose inequality. Notice how the slice passes just below the event
    horizon in Portion 1. Therefore, the apparent horizon is located where the
    initial data intersects the marginally outer trapped 2-spheres in the
    second Schwarzschild region, Portion 3.}
  \label{CounterFig}
\end{figure}

First, the matching of Portion 1 and 2. Portion 1 is matched along the
original surface used to obtain it. Portion 2 is matched along the surface
spanned by $\chi=\chi_0$. The junction conditions are satisfied when
\bean
R_0=a_m \sin\chi_0 \label{1-2a} \\
M=\frac{1}{2}a_m \sin^3\chi_0 \label{1-2b}
\eean
This is the same matching as in a generalized OS spacetime with
$\chi_0>\frac{\pi}{2}$.

Next, the matching of Portion 2 and 3. Portion 2 is matched along the surface
spanned by $\chi=\chi_1$. Portion 3 is matched along the surface with geodesics
passing at $R_1$ at $t=0$. The junction conditions for this matching are
satisfied when
\bean
R_1=a_m \sin\chi_1 \label{2-3a} \\
M_1=\frac{1}{2}a_m \sin^3\chi_1 \label{2-3b}
\eean

Finally, the matching of Portion 3 and 4. Portion 3 is matched along the
surface with geodesics passing at $R_2$ at $t=0$. Portion 4 is matched along
the surface spanned by $\chi=\chi_2$. The junction conditions are satisfied
when

\bean
R_2=a^\prime_m \sin\chi_2 \label{3-4a} \\
M_1=\frac{1}{2}a^\prime_m \sin^3\chi_2 \label{3-4b}
\eean

A choice of all these parameters satisfying the junction conditions uniquely
determines the spacetime. However, there is freedom in choosing the parameters
leading to a 4-parameter family of spacetimes constructed in this way: Choose
any $M$ and any $\chi_0\in(\frac{\pi}{2},\pi)$. This uniquely determines
$R_0$ and $a_m$ by (\ref{1-2a}),(\ref{1-2b}). A choice of any
$\chi_1\in[\frac{\pi}{2},\chi_0)$ now uniquely determines
$M_1$ and $R_1$ by (\ref{2-3a}),(\ref{2-3b}). Finally a choice of any $R_2>R_1$
uniquely determines $a^\prime_m$ and $\chi_2$ by (\ref{3-4a}),(\ref{3-4b}).

\medskip

Certain members of this 4-parameter family of spacetimes contain slices that
are counterexamples to the apparent horizon Penrose inequality as will now be
demonstrated.

Given any $M$ and any $\chi_0\in(\frac{\pi}{2},\pi)$ choose $\chi_1$ in its
allowed range, i.e.\ $\chi_1\in[\frac{\pi}{2},\chi_0)$.
(In Fig.\ \ref{CounterFig}, $\chi_1=\frac{\pi}{2}$.)\,
This determines $M_1$ and $R_1$. It remains to choose $R_2$ as long as
$R_2>R_1$. Later it will turn out that a counterexample is obtained as long as
$R_2$ is large enough, thus such a choice can then be made.

The initial data is obtained by taking the following slice: In Portion 1,
part of a Schwarzschild spacetime, the slice is at constant
Kruskal-Szekeres coordinate T (i.e.\ it is a horizontal 
slice in a Kruskal diagram). It is chosen so as to pass just below the event
horizon in Portion 1 (see Fig. \ref{CounterFig}). In
Portion 2, part of a RW spacetime, the slice continues at fixed $\eta$. In
Portion 3, part of another Schwarzschild spacetime, the slice continues, again,
at fixed T. In Portion 3, the slice intersects the 3-surface corresponding to
the event horizon of the Schwarzschild spacetime from which this portion
originates. This intersection is a marginally outer trapped 2-sphere.
Finally in Portion 4, part of a RW spacetime, the
slice continues at fixed $\eta$.

In order to guarantee that this slice contains such a marginally outer trapped
2-sphere in Portion 3, $R_2$ must be chosen in such a way that
Portion 4 will start ``further back''. This is satisfied if $R_2$ is large 
enough; thus, a suitable value of $R_2$ is now chosen. 

In this initial data, the apparent horizon, the outermost marginally outer
trapped surface, is just that marginally outer trapped 2-sphere in Portion 3,
as no surface outside it is marginally outer trapped. The area of the apparent
horizon, in this case, is therefore given by
\bean \label{area1}
A=4\pi(2M_1)^2
\eean
whereas, in contrast, the total (ADM) mass of this initial
data is $M$, as this is evaluated in the asymptotically flat region.
As $\chi_1\in[\frac{\pi}{2},\chi_0)$ then $\sin\chi_1 > \sin\chi_0$
and then from (\ref{1-2b}),(\ref{2-3b}) it follows that
\bean \label{greaterM}
M_1 > M
\eean
Finally, (\ref{area1}),(\ref{greaterM}) imply:

\bean
M < \sqrt{A/16\pi}
\eean

This is an asymptotically flat spacetime satisfying the dominant energy 
condition. It is spherically symmetric with a spherically symmetric slice
yielding initial data that violates the apparent horizon Penrose inequality.
Furthermore, as $\chi_0$ can be chosen to be arbitrarily close to $\pi$,
$\sin\chi_0$ can be made as small as one wishes and therefore $A$ can be
made as large as one wishes. Thus, for all $\eps>0$ there
exists a spacetime and a slice as described above producing initial data 
with $M<\eps\sqrt{A/16\pi}$ with $A$ the area of the apparent horizon,
i.e.\ the inequality can be violated to an arbitrarily large extent.

\bigskip
The fact that $M_1$ is bigger than $M$, so that this construction can work as
a counterexample to the apparent horizon Penrose inequality, is due to specific
features of Portion 2. As $\Sigma_{(l)}$ in that portion is foliated by
marginally outer trapped 2-spheres, one can define $A(\eta)$ to be the area of
these 2-spheres. Using (\ref{cyclic1}),(\ref{ChiOfEta}), this is found to be
\bean
A(\eta) = 4\pi \, a^2(\eta) \sin^2 \chi_l(\eta) = \frac{\pi}{2} a_m^2
(1-\cos\eta)^3
\eean
Thus, this area increases during the expansion ($0\leq\eta < \pi $) of the RW
portion and decreases during its collapse ($\pi<\eta\leq 2\pi $). As the
matching of Portions 1 and 2 and then that of Portions 2 and 3 is such
that $\Sigma_{(l)}$ in Portion 2 is limited to an expanding RW 
region,\footnote{In a normal OS spacetime, this is not the case,
as there $\Sigma_{(l)}$ is located in a collapsing RW region.
This is the reason why Portion 1 and 2 are chosen so that the matching
between them is like that in a generalized OS spacetime.}\, the area of
marginally outer trapped 2-spheres in this portion is growing. It is this
bigger area that implies a bigger mass parameter, $M_1$, in
Portion 3.\footnote{Since $\Sigma_{(l)}$
is a timelike surface, the initial data slice cannot be modified to intersect
it, so as to produce a counterexample with only two portions. This is why
the second Schwarzschild portion, Portion 3, is required. Portion 4 is
taken only to ``close the cap'' on the other asymptotically flat region,
i.e.\ obtain a spacetime with only one asymptotically flat end, in Portion 1.}


\section{Quasi-local constructions} \label{QuasiLocal}

Before remarking on various versions of the Penrose inequality, it is important
to discuss an issue raised by $\Sigma_{(l)}$ being a timelike surface in all
the constructions above.

In recent years certain quasi-local constructions were suggested as either 
useful in applications pertaining to the dynamics of black holes or
even as candidates to replace event horizons altogether as their boundaries.
Recently, Ashtekar and Krishnan \cite{AshKrish1} defined and discussed
dynamical horizons, building on previous formulations of isolated horizons
\cite{Ash&1} and Hayward's \cite{Hay94} notion of trapping horizons. The
following definitions follow the works where they were originally defined.

A {\it dynamical horizon} is a smooth, three-dimensional, spacelike
sub-manifold that can be foliated by a family of closed 2-surfaces with the
expansion of future-directed null geodesics normal to these 2-surfaces
vanishing in one direction and strictly negative in the
other.\footnote{Ashtekar and Krishnan do
not rely on asymptotic flatness. In the context of the discussion here, a
dynamical horizon would be foliated by 2-surfaces that are inner trapped and
marginally outer trapped, with that expansion vanishing everywhere, or by
2-surfaces that are outer trapped and marginally inner trapped, with that
expansion vanishing everywhere.}

A {\it future outer trapping horizon} is the closure of a 3-surface foliated by
2-surfaces on which: The expansion of future-directed null geodesics normal to
the 2-surface in one direction vanishes, denote it as $\theta_{(+)}$.
The expansion in the other direction, $\theta_{(-)}$, is strictly negative,
and finally $\mathcal{L}_-\theta_{+}$ is also strictly negative.

In asymptotically flat spacetimes both dynamical horizons\footnote{With $l^a$
directed towards the asymptotically flat end.} and future outer trapping
horizons\footnote{With the '+' direction, in this case, pointing outwards,
i.e.\ to the asymptotically flat end.} are 3-surfaces that are foliated
by marginally outer trapped surfaces. Moreover, dynamical horizons are
spacelike by definition, and future outer trapping horizons must be
spacelike or null \cite{AshKrish1}. In contrast, in all of the spacetimes
constructed here, the 3-surfaces foliated by marginally outer trapped 2-spheres
are never spacelike. They are null in the Schwarzschild portions and are
timelike where there is matter and dynamics, i.e.\ in the RW portions.
This raises the issue of whether dynamical horizons or future outer trapping
horizons are associated, generically, with gravitational collapse. 

Given a 3-surface foliated by marginally outer trapped surfaces, Ashtekar
and Krishnan \cite{AshKrish1} show that the question whether the 3-surface is
timelike or spacelike is directly related to $\mathcal{L}_n\theta_{(l)}$,
the derivative of the expansion of future, outgoing, null geodesics
normal to the 2-surfaces in the direction of ingoing ones. If the 3-surface
is not null and if $\mathcal{L}_n\theta_{(l)}$ is non-zero, then they show,
using the Raychaudhuri equation, that the 3-surface is spacelike if
$\mathcal{L}_n\theta_{(l)}$ is negative and it is timelike if
$\mathcal{L}_n\theta_{(l)}$ is positive. Thus, the issue translates to
whether $\mathcal{L}_n\theta_{(l)}$ is generically positive or negative.

Ashtekar and Krishnan \cite{AshKrish1} provide arguments for
$\mathcal{L}_n\theta_{(l)}$ to be generically negative, and similarly Hayward 
\cite{Hay01} provides an argument for future outer trapping horizons to be
generic. However, none of these arguments seem satisfactory.

As a concrete example of dynamical horizons Ashtekar and Krishnan 
\cite{AshKrish1} provide the Vaidya spacetimes. The OS spacetimes provide
a concrete example where a 3-surface, foliated by marginally outer trapped
surfaces, is timelike. Both the Vaidya and OS spacetimes are spherically 
symmetric, and in spherical symmetry the issue of whether such a 3-surface
is spacelike or timelike, can be further explored via the Newman--Penrose 
formalism \cite{NewmanPenrose}.\footnote{The
sign conventions used in \cite{NewmanPenrose} differ substantially from
Wald \cite{WaldGR} and are adjusted for as follows: When dealing with symbols
defined in the work of Newman and Penrose, the expressions appearing here will
have the original signs, as used by Newman and Penrose. However, once these
expressions are put in more conventional form, the conventions of
Wald \cite{WaldGR} shall be imposed. In particular, once an expression is
rewritten in terms of components of the Weyl tensor and scalar curvature,
these are given with the sign conventions of Wald \cite{WaldGR}.}

The work of Newman and Penrose involves a spinor formalism for general
relativity. In \cite{NewmanPenrose} Newman and Penrose take two null vectors,
$l^a$, $n^a$ with normalization\footnote{As the signature in 
\cite{NewmanPenrose} is +2 they actually have $l^a n_a = 1$.}
$l^a n_a = -1$. With respect to some scalar quantity $\phi$ they define in
(2.12) $\Delta\phi = n^a \nabla_a \phi$ and the spin coefficient $\rho$ is a
function defined in (4.1a) such that $\rho=\frac{1}{2}\theta_{(l)}$. Thus, up 
to a numerical factor, $\Delta\rho$ is $\mathcal{L}_n\theta_{(l)}$, hence
equation (4.2q) is useful.

In the spherical symmetric case and on the marginally outer trapped surface,
where $\rho=0$, most terms in (4.2q) vanish and it reduces to
\bean \label{NP4.2q}
\Delta\rho = -\psi_2 - 2\Lambda
\eean

As $\psi_2$ is, up to a numerical factor, a component of the Weyl tensor 
and $\Lambda$ is, up to a numerical factor, $R$, the scalar curvature, this
becomes
\bean
\mathcal{L}_n\theta_{(l)} = C_{abcd}\,l^a n^b\,l^c\, n^d
+ \frac{R}{6} \label{NPeq}
\eean
In this form it is easier to see how $\mathcal{L}_n\theta_{(l)}$ is positive in
one case and is negative in the other.

The Vaidya spacetime is one that contains a null fluid. As a result of this
form of stress energy, the scalar curvature in this case vanishes. Hence, in
the ingoing Eddington-Finkelstein coordinates used by Ashtekar and Krishnan
(\ref{NPeq}) becomes
\bean
\mathcal{L}_n\theta_{(l)} = C_{abcd}\,l^a n^b\,l^c\, n^d = C_{vrvr} = -
\frac{1}{r^2}
\eean

This shows, in agreement with a direct evaluation of the LHS as in
\cite{AshKrish1},\footnote{In their paper, Ashtekar and Krishnan take
$l^a n_a = -2$ hence what they evaluate as $\mathcal{L}_n\theta_{(l)}$ is
twice the value obtained here.} that in the Vaidya spacetime
$\mathcal{L}_n\theta_{(l)}$ is negative, and the 3-surface foliated by
marginally outer trapped 2-surfaces is indeed spacelike where the
stress-energy is non-vanishing \cite{AshKrish1}.

In the case of a RW spacetime, the non-vacuum portion of an OS spacetime, the
Weyl tensor vanishes.\footnote{A RW spacetime is conformally equivalent to a
spacetime with constant curvature where the Weyl tensor is known to vanish
identically, and since the Weyl tensor is conformally invariant, it must
vanish for all RW spacetimes.}\, Thus, in this case, (\ref{NPeq}) becomes
\bean
\mathcal{L}_n\theta_{(l)} = \frac{R}{6}
\eean

In the RW spacetime $R=8\pi(\rho - 3P)$ so that in the case of dust $R>0$.
Consequently $\mathcal{L}_n\theta_{(l)}>0$, implying that $\Sigma_{(l)}$ is
timelike, in agreement with the result shown in the appendix.

Normal matter satisfies $0 \leq P \leq \frac{1}{3}\rho$ and hence the 
contribution from the scalar curvature should generically be positive.
As for the Weyl tensor, it does not seem clear what sign this component should
generically have. Nor is it clear what the relation is, if any, between the
Weyl tensor and the scalar curvature---which one generically
dominates?\footnote{I. Booth (private communication) has shown that
(\ref{NP4.2q}) can be transformed to an expression that is more useful in
determining the sign of $\mathcal{L}_n\theta_{(l)}$. In spherical symmetry,
using (4.2l) of \cite{NewmanPenrose}, (\ref{NP4.2q}) becomes:
\bea
\mathcal{L}_n\theta_{(l)} = -2(3\Lambda+\phi_{11}) - \frac{1}{r^2}  \\
 = 8\pi T_{ab} l^a n^b - \frac{1}{r^2} \\
 = 4\pi (\rho-P_r) - \frac{1}{r^2} 
\eea
where $r$ is the areal radius of the marginally outer trapped surface,
$\rho=T_{ab}t^a t^b$ is the energy density, and
$P_r=T_{ab} r^a r^b$ is the radial pressure. (Here $t^a$ and $r^a$ 
are orthonormal vectors that are orthogonal to the 2-spheres.)
The dominant energy condition implies that $\rho-P_r$ is non-negative.
Thus, the sign of $\mathcal{L}_n\theta_{(l)}$ now depends on two terms: a
stress-energy term that tends to make $\mathcal{L}_n\theta_{(l)}$ positive 
and a term proportional to the inverse of the area of the horizon that tends
to make $\mathcal{L}_n\theta_{(l)}$ negative. Hence, if the density of matter
is large compared with the inverse of the area of the horizon, and if the
radial pressure is sufficiently small, then $\mathcal{L}_n\theta_{(l)}$ will be
positive and the 3-surface timelike. (This is what happens in RW spacetimes
with small enough pressure.) If, however, the converse is true then
$\mathcal{L}_n\theta_{(l)}$ will be negative and the 3-surface will be
spacelike (as is the case in Vaidya spacetimes, since with null fluid the
stress-energy term vanishes).}\, Note however, that the above
results hold only in the spherically symmetric
case. In a non-spherically-symmetric spacetime (4.2q) in \cite{NewmanPenrose}
will have additional non-vanishing terms and it is therefore much harder to
make a precise statement in the general case. Furthermore, in the
non-spherically-symmetric case there is no reason to expect 
$\mathcal{L}_n\theta_{(l)}$ to have a constant sign over the entire
marginally outer trapped surface.


\section{Discussion} \label{Discussion}

Though the counterexample above shows that the apparent horizon Penrose
inequality does not hold, it does not contradict the Penrose inequality and it
therefore has no effect on the status of the cosmic censorship conjecture.

\bigskip

In the spherically symmetric case, Malec and \'O Murchadha \cite{MalecMurcha1}
considered another version of the Penrose inequality. Given a spherically
symmetric initial data surface, one can not only locate the apparent
horizon in it, but also the past apparent horizon (the apparent horizon
with respect to past-directed null geodesics).\footnote{The past apparent
horizon is just the apparent horizon in the initial data obtained by time
reversing the given one.}\, Define the {\it outermost horizon} to be the 
outermost of the future and past apparent horizons. 
Malec and \'O Murchadha's version of the Penrose inequality, which in the
present work will be referred to as 
{\bf the outermost horizon Penrose inequality}, is that given a {\it
spherically symmetric} initial data satisfying the dominant energy 
condition\footnote{In fact, in \cite{MalecMurcha1} the dominant energy
condition is only required to hold outside the outermost horizon.} then
\bean
M \, \geq \, \sqrt{A_{o.h}/16\pi}
\eean

\noindent
with $M$ the total mass and $A_{o.h}$ the area of the outermost
horizon. 

In \cite{MalecMurcha1}, Malec and \'O Murchadha proved the outermost horizon
Penrose inequality for maximal slices.\footnote{Unfortunately, the wording of
the statement of the theorem in \cite{MalecMurcha1} can be interpreted as
asserting that the apparent horizon Penrose inequality holds in spherical
symmetry. In fact in Szabados \cite{Szabad} the results of \cite{MalecMurcha1}
appear to have been interpreted in such a way.}\, This was later proven
without requiring maximal slices by Iriondo, Malec, and \'O Murchadha
\cite{Malec&96} and independently by Hayward \cite{Hay96,Hay98}.

\bigskip

The proofs of the outermost horizon Penrose inequality may suggest that some
version of it may hold for initial data lacking spherical symmetry. In such
initial data the (future) apparent horizon may not lie completely outside
or completely inside the past apparent horizon.
The future and past apparent horizons may then intersect in a
complicated manner, in which case care must be taken in defining the
outermost horizon in initial data without spherical symmetry.
One might consider the union of the (future)
trapped region and the past trapped region and define the outermost horizon
to be the outermost boundary of this union. This, however, will yield
a surface that in general need not be smooth.

Another approach to generalizing the formulation of the outermost horizon 
Penrose inequality to non-spherically-symmetric initial data might be 
to restrict consideration to the
case where the apparent horizon lies completely outside the past apparent
horizon, i.e., to formulate the outermost horizon Penrose inequality only in 
the case where there are no past outer trapped surfaces outside the
apparent horizon.\footnote{In spherical symmetry this is not a restriction
since, in the context of the outermost horizon Penrose inequality, if this
condition fails, it will hold for the time reverse of the initial 
data.}\, However, it is possible that a spacetime like the one presented in
this work can serve as a counterexample for this case as well
using a highly non-spherically-symmetric slice, similar to that
used in \cite{WaldIyer}. In this case coming from the asymptotically flat end,
the north pole of the 2-spheres might be taken to get very close to the white
hole singularity, while the south pole of the 2-spheres does not. Further
inside, the slice will approach the spherically symmetric
slice shown in Fig. \ref{CounterFig}. In this way
it may be possible to obtain a slice that does not contain any past outer
trapped surfaces outside the apparent horizon and in addition, if the
apparent horizon is located in the first RW portion or the second
Schwarzschild portion, its area might be large enough so as to obtain a
counterexample.

\bigskip

It is important to realize that the outermost horizon Penrose inequality
actually implies the Penrose inequality in spherical symmetry. In this case
$M\geq \sqrt{A_{o.m.}/16\pi}$ with $A_{o.m.}$ the area of the outermost
horizon. However, the outermost horizon either coincides with, or lies outside
of the apparent horizon. In either case it follows
immediately that $M \geq \sqrt{\mathcal{A}/16\pi}$ with $\mathcal{A}$ the
minimum area required to enclose the apparent horizon. Thus, in spherical
symmetry the ``true'' Penrose inequality has been proven, and it would be of
great interest to extend this to the general case.

\bigskip


\noindent
{\bf Acknowledgments}

\medskip

It is a pleasure to thank Robert M. Wald for guidance and a lot of useful
advice along the way, not to mention suggesting this project in the first
place. I would also like to thank Mike Seifert, Stefan Hollands, and
Piyush Kumar for many useful discussions and
suggestions.
This research was supported in part by NSF grant PHY 00-90138 to
the University of Chicago.


\appendix\section{The type of $\Sigma_{(l)}$ and $\Sigma_{(n)}$}

The line element of a Robertson--Walker spacetime is given by
\bean \label{RWmetric}
d s^2=-d \tau^2 + a^2(\tau)\Big(dr^2 + f(r)^2 d\Omega^2 \Big)
\eean

\noindent
with $f(r)$, depending on the spatial geometry, given by
\bea
\left\{ \begin{array}{l@{\quad\quad\quad}l}
\sin r, & \text{closed (}k=1\text{) geometry} \\
r, & \text{flat (}k=0\text{) geometry} \\
\sinh r, & \text{open (}k=-1\text{) geometry} \\ 
\end{array} \right.
\eea

The spacetime is assumed to have a stress energy tensor of a perfect fluid form
with an equation of state $P=\omega \rho$ with $\rho>0$.

The Friedman equations, the field equations applied to such spacetimes are
given \cite{WaldGR} by
\bean
3\dot{a}^2/a^2 = 8\pi\rho - 3k/a^2 \label{Friedman1} \\
3\ddot{a}/a = -4\pi(\rho + 3P) \label{Friedman2}
\eean

The 3-surfaces $\Sigma_{(l)}$ and $\Sigma_{(n)}$ are given by $r=r(\tau)$ where
the expansion of the relevant geodesics vanish. The radial tangents to this 
surface are found to be\footnote{The '$+$' sign is
for the south trapped case, i.e.\ increasing radial coordinate, and the '$-$'
sign for the north trapped case, i.e.\ decreasing radial coordinate.} 

\bean \label{tangent}
u^a = \Big(\frac{\partial}{\partial \tau}\Big)^a \pm 
\frac{\ddot{a}}{\dot{a}^2+k} \Big(\frac{\partial}{\partial r}\Big)^a
\eean

Hence 
\bean
u^a u_a = \Big(\frac{\ddot{a}a}{\dot{a}^2+k}\Big)^2-1 =
\Big(\frac{\rho + 3P}{2\rho}\Big)^2-1
\eean

\noindent
where in the last equality, (\ref{Friedman1}) and (\ref{Friedman2}) have been
used.

It follows that $u^a$, and therefore the 3-surface, is
\bea
\left\{ \begin{array}{l@{\quad\quad\quad}l}
\text{timelike,} & -1< \omega < \frac{1}{3} \\ 
\text{null,} & \omega = \frac{1}{3} , -1 \\
\text{spacelike,} & \omega > \frac{1}{3} \text{ or } \omega < -1
\end{array} \right.
\eea
This is independent of the spatial geometry of the RW model.


\end{document}